\documentclass[11pt]{article}

\usepackage{latexsym,ifthen,epsfig}
\usepackage[latin1]{inputenc}
\usepackage{amsmath}

\newcommand{\qed}{\hfill $\Box$}

\newcommand{\0}{\text{ has a co-join to }}
\newcommand{\1}{\text{ has a join to }}

\newtheorem{theo}{Theorem}
\newtheorem{lemm}{Lemma}
\newtheorem{coro}{Corollary}

\newtheorem{clai}{Claim}[section]
\newtheorem{prop}{Proposition}
\newtheorem{obse}{Observation}

\title{Maximum Weight Independent Sets in Odd-Hole-Free Graphs Without Dart or Without Bull} 

\author{
Andreas Brandst\"adt\thanks{Institut f\"ur Informatik, Universit\"at Rostock, 
D-18051 Rostock, Germany. E-mail: ab@informatik.uni-rostock.de} 
\and
Raffaele Mosca\thanks{Dipartimento di Economia, Universit\'a degli Studi ``G. d'Annunzio'', Pescara 65121, Italy. 
E-mail: r.mosca@unich.it}}
\date{}
\begin{document}

\maketitle

\begin{abstract} 
The Maximum Weight Independent Set (MWIS) Problem on graphs with vertex weights asks for a set of pairwise nonadjacent vertices of maximum total weight. Being one of the most investigated and most important problems on graphs, it is well known to be NP-complete and hard to approximate. 
The complexity of MWIS is open for hole-free graphs (i.e., graphs without induced subgraphs isomorphic to a chordless cycle of length at least five). By applying clique separator decomposition as well as modular decomposition, we obtain polynomial time solutions of MWIS for odd-hole- and dart-free graphs as well as for odd-hole- and bull-free graphs ({\em dart} and {\em bull} have five vertices, say $a,b,c,d,e$, and dart has edges $ab,ac,ad,bd,cd,de$, while bull has edges 
$ab,bc,cd,be,ce$). If the graphs are hole-free instead of odd-hole-free then stronger structural results and better time bounds are obtained.
\end{abstract}


\section{Introduction}

The Maximum Weight Independent Set (MWIS) Problem on a given finite undirected simple graph with vertex weights asks for a set of pairwise nonadjacent vertices of maximum total weight. Being one of the most investigated and most important problems on graphs, it is well known to be NP-complete and hard to approximate. It is solvable in polynomial time on various graph classes while it remains NP-complete on some others. Its complexity is open for hole-free graphs (i.e., graphs without induced subgraphs isomorphic to a chordless cycle of length at least five). Recently, the following subclasses of hole-free graphs were studied: 
\begin{itemize} 
\item[(i)] hole-free graphs without induced diamond and in general without induced paraglider \cite{BraGiaMaf2012}; 
\item[(ii)] hole-free graphs without induced co-chair \cite{BraGia2012};
\item[(iii)] hole-free graphs without induced dart \cite{BasChaKar2012}.  
\end{itemize} 

In \cite{BraGiaMaf2012}, it was shown that the atoms of the graphs in the class (i) are weakly chordal or specific graphs, and in \cite{BraGia2012}, 
it was shown that the prime atoms of the graphs in the class (ii) are nearly weakly chordal, from which polynomial time algorithms for MWIS on these graphs follow.  

Using the approach of \cite{BraGia2012,BraGiaMaf2012}, Basavaraju, Chandran and Karthick in \cite{BasChaKar2012} showed that the MWIS problem can be solved in polynomial time for hole- and dart-free graphs by reducing it first to hole-, dart-, and gem-free graphs (which are hole- and paraglider-free). We extend previous results and show:
\begin{itemize}
\item[(i)] odd-hole- and dart-free graphs are nearly perfect, and (hole, dart)-free atoms are nearly weakly chordal. 
\item[(ii)] odd-hole- and bull-free prime graphs are nearly perfect, and (hole, bull)-free prime graphs are nearly weakly chordal. 
\item[(iii)] hole- and dart-free graphs as well as hole- and bull-free graphs have nice structure properties; MWIS for hole- and bull-free graphs can be solved in time ${\cal O}(n^5)$. 
\item[(iv)] MWIS for $P_5$- and bull-free graphs can be solved in time ${\cal O}(n m)$. 
\end{itemize}

The results in (i) and (ii) are based on the Strong Perfect Graph Theorem and imply that the MWIS problem is solvable in polynomial time for odd-hole- and dart-free graphs as well as for odd-hole- and bull-free graphs. 

Actually, Lemma 1 in \cite{BraGia2012} implies that also (odd-hole,co-chair)-free prime atoms are nearly perfect (which implies polynomial time for the MWIS problem on (odd-hole,co-chair)-free graphs). This fact was not explicitely mentioned in \cite{BraGia2012}. 

\section{Some Basic Notions and Results}  

\subsection{Basic Notions}  

For any missing notation or reference let us refer to \cite{BraLeSpi1999}. Let $G = (V,E)$ be a finite undirected graph which is simple (i.e., without self-loops and multiple edges) with vertex set $V$ and edge set $E$. 
Let $|V|=n$ and $|E|=m$. 
For a vertex $v \in V$, let $N(v):=\{u \mid uv \in E\}$ denote the {\em open neighborhood} of $v$ and let $N[v]:=N(v) \cup \{v\}$ denote the {\em closed neighborhood} of $v$. We also say that for vertices $u,v \in V$, $u$ and $v$ {\em see each other} ({\em miss each other}, respectively) if $uv \in E$ ($uv \notin E$, respectively). 
Let $A(v):=V \setminus N[v]$ denote the {\em anti-neighborhood} of $v$.  

For any vertex set $U \subseteq V$, let $N(U) := \bigcup_{u \in U} N(u) \setminus U$ and $N[U]:=N(U) \cup U$ as well as $A(U):=V \setminus N[U]$.
For any nonempty vertex subset $U \subseteq V$ with $A(U) \neq \emptyset$, let $U^+ := N(U) \cap N(A(U))$ (the set of {\em contact vertices} of $U$ and $A(U)$).

For any vertex set $U \subseteq V$ let $G[U]$ be the subgraph of $G$ induced by $U$. Let $\overline{G}=(V,\overline{E})$ denote the {\em complement graph of $G$}, also denoted as co-$G$. Pairs $xy \in \overline{E}$ are also called {\em co-edges} of $G$. 

For any disjoint vertex sets $U,W$ of $V$, let us say that $U$ has a {\em join} (a {\em co-join}, respectively) to $W$ if each vertex of $U$ sees each vertex of $W$ (misses each vertex of $W$, respectively). If $U = \{u\}$ and $U$ has a join to $W$, then let us say that $u$ {\em dominates} $W$ or is {\em universal for $W$}.

\begin{figure}
  \begin{center}
    \epsfig{file=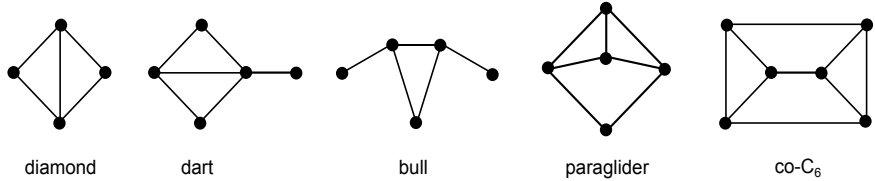}
    \caption{diamond, dart, bull, paraglider, and co-$C_6$}
    \label{dart}
  \end{center}
\end{figure}

$P_k$ is the induced path with $k$ vertices and $k-1$ edges. $C_k$ is the induced cycle with $k$ vertices and $k$ edges. A {\em hole} is $C_k$ with $k \geq 5$. An 
{\em odd hole} is $C_{2k+1}$ with $k \geq 2$. An {\em anti-hole} is the complement graph of a hole. 
A {\em diamond} (or $K_4-e$) is formed by vertices $a,b,c,d$, and edges $ab,ac,ad,bd,cd$. A {\em gem} is a one-vertex extension of a diamond, which can be obtained by adding  a dominating vertex to a $P_4$.
A {\em dart} has five vertices $a,b,c,d,e$, and edges $ab,ac,ad,bd,cd,de$, i.e., it consists of a diamond plus a degree-one vertex being adjacent to one of the degree-3 vertices of the diamond. A {\em bull} has five vertices $a,b,c,d,e$ and edges $ab,bc,cd,be,ce$. See Figure \ref{dart} for most of these specific graphs. 

For any graph $H$, let us say that $G$ is {\em $H$-free} if $G$ contains no induced subgraph isomorphic to $H$. A class of graphs is {\em hereditary} if it is closed under taking induced subgraphs. A graph is {\em chordal} it it is hole-free and $C_4$-free. A graph is {\em weakly chordal} if it is hole- and anti-hole-free.
{\em Perfect graphs} play a crucial role in algorithmic graph theory, and the Strong Perfect Graph Theorem (see Theorem \ref{SPGT}) characterized them in terms of forbidden odd holes and odd anti-holes. It is well known that chordal graphs are weakly chordal, and weakly chordal graphs are perfect (see e.g. \cite{BraLeSpi1999} for this and the many facets of these graphs). 

An {\em independent set} of $G$ is a subset of pairwise non-adjacent vertices of $G$. A $clique$ of $G$ is a set of pairwise adjacent vertices of $G$.

\subsection{Techniques for the Maximum Weight Independent Set Problem}
 
The Maximum Independent Set (MIS) Problem asks for an independent vertex set of maximum cardinality. If $w$ is a real-valued function on $V$ then the Maximum Weight Independent Set (MWIS) Problem asks for an independent set of maximum total weight. Let $\alpha_w(G)$ denote the maximum weight of an independent vertex set in $G$.

A subset $U \subseteq V$ is a $cutset$ (or $separator$) in $G$ if $G[V \setminus U]$ has more connected components than $G$. A {\em clique cutset} is a cutset which is a clique. An $atom$ in $G$ is an induced subgraph of $G$ without clique cutset. More generally, a graph is an $atom$ if it has no clique cutset. A famous divide-and-conquer approach by using clique separators is described in \cite{Tarjan1985,White1984}. A consequence is the following:

\begin{theo}[\cite{Tarjan1985,White1984}]\label{MWISatomred}
If for a hereditary graph class ${\cal C}$, the MWIS problem is solvable in polynomial time for the atoms of ${\cal C}$ then the MWIS problem is solvable in polynomial time on graph class ${\cal C}$.
\end{theo}

Let $\Pi$ denote a (hereditary) graph property. A graph $G=(V,E)$ is {\em nearly $\Pi$} if for all $v \in V$, the subgraph $G[A(v)]$ has property $\Pi$. For short, we say that for every vertex, the anti-neighborhood has property $\Pi$. For example, $G$ is {\em nearly weakly chordal} if the anti-neighborhood of every vertex is weakly chordal. Obviously the following holds:

\begin{obse}
$\alpha_w(G) = \max\{w(v) + \alpha_w(G[A(v)]) \mid v \in V\}$
\end{obse}

Thus, we obtain: 

\begin{coro}\label{nearlyPiMWIS}
Whenever MWIS can be solved in time $T$ on a hereditary graph class with property $\Pi$, it can be solved in time $|V| \cdot T$ on nearly $\Pi$ graphs. 
\end{coro}

For example, since the MWIS problem can be solved in time ${\cal O}(n^4)$ for weakly chordal graphs \cite{SpiSri1995}, it can be solved in time ${\cal O}(n^5)$ for nearly weakly chordal graphs.

In Section \ref{holebullfr} we make use of modular decomposition. We say that a vertex $z$ {\em distinguishes two vertices $x,y$} if $z$ sees $x$ and misses $y$. A subset $U$ of vertices is a {\em module in $G$} if no vertex $z \in V \setminus U$ distinguishes two vertices $x,y \in U$. A module is {\em trivial} if it is either the empty set or $V$ or one-elementary. A graph is {\em prime} if all its modules are trivial. We will use primality for solving MWIS on ((odd-)hole,bull)-free graphs. It is well known that MWIS can be solved in time $T$ bottom-up along the modular decomposition tree for a hereditary graph class ${\cal C}$ if MWIS can be solved in time $T$ for the prime graphs of ${\cal C}$; the modular decomposition tree of a given graph can be determined in linear time \cite{McCSpi1999}. 
    
\begin{theo}\label{MWISprimered}
If for a hereditary graph class ${\cal C}$, the MWIS problem is solvable in time $T$, $T \ge m$, for the prime graphs of ${\cal C}$ then the MWIS problem is solvable in time ${\cal O}(T)$ on graph class ${\cal C}$.
\end{theo} 

\subsection{Basic Results on Some Classes of Perfect and Other Graphs}

Subsequently, we use the following results and facts:

\begin{theo}[Strong Perfect Graph Theorem \cite{ChuRobSeyTho2006}]\label{SPGT}
A graph is perfect if and only if it is odd-hole-free and odd-anti-hole-free.
\end{theo}

\begin{theo}\label{basicrecprop}
The following graph classes can be recognized in polynomial time:
\begin{enumerate}
\item[$(i)$] weakly chordal graphs $\cite{BerBorHeg2000,HaySpiSri2007,SpiSri1995}$
\item[$(ii)$] hole-free graphs $\cite{EscSri1995,Spinrad1991}$ 
\item[$(iii)$] perfect graphs $\cite{CorLiuVus2003}$.
\end{enumerate}
\end{theo}
 
Recognition of odd-hole-free graphs is open, though recognition of odd-hole-free graphs with cliques of bounded size can be done in polynomial time \cite{ConCorLiuVusZam}.
 
\begin{theo}\label{basicMWISprop}
The MWIS problem for can be solved in polynomial time for the following graph classes:
\begin{enumerate}
\item[$(i)$] weakly chordal graphs $\cite{SpiSri1995}$; the time bound is ${\cal O}(n^4)$.
\item[$(ii)$] perfect graphs $\cite{GroLovSch1984}$.
\item[$(iii)$] hole-free graphs with no banner $\cite{BraKleLozMos2010}$, with no paraglider $\cite{BraGiaMaf2012}$, with no co-chair $\cite{BraGia2012}$ and with no dart \cite{BasChaKar2012}.
\end{enumerate}
\end{theo} 

The MWIS problem for hole-free graphs (and for odd-hole-free graphs) is open; it is open even for the subclass of ($P_5,C_5$)-free graphs. 

Finally let us mention a recent paper \cite{Chudn2011} introducing many structural properties for bull-free graphs (see also \cite{DeFMafPor1997} for perfect bull-free graphs) and the papers \cite{BraHoaLe2003,DeSSas1993} focussing on efficiently solving MWIS for (bull,chair)-free graphs. 

\section{Structure and MWIS for (Odd-)Hole- and Dart-Free Graphs}

In this section, we show that odd-hole- and dart-free graphs are nearly perfect and that hole- and dart-free atoms are nearly weakly chordal. This is the main result of the section and implies polynomial time for MWIS on both graph classes (with better time bound for hole- and dart-free graphs). We first collect some properties of dart-free graphs and deal with (odd-hole,dart)-free graphs.. 

\subsection{Structure and MWIS for (Odd-Hole,Dart)-Free Graphs}

\begin{prop}\label{prop:1}
Let $G = (V,E)$ be a dart-free graph and let $U \subset V$. If $u \in U^+$ and vertices $a,b,c$ induce a $P_3$ in $U$ then $u$ does not see all three vertices 
$a,b,c$. 
\end{prop}

\noindent
{\bf Proof.} Otherwise, since $u \in U^+$ sees a vertex $v \in A(U)$ (recall that $A(U) \neq \emptyset$), such vertices $a,b,c$ together with $u$ and $v$ would induce a dart. \qed

\medskip

Subsequently, when dealing with cycles of length $k$ as well as their complements, index arithmetic is done modulo $k$.

\begin{lemm}\label{lemm:1}
Let $G = (V,E)$ be a dart-free graph containing an anti-hole $\overline{C_k}$, $k \geq 7$, say $H$, with vertices $v_1,\ldots,v_k$ and co-edges $v_iv_{i+1}$ for $i = 1,\ldots,k$, such that $H^+ \neq \emptyset$, and let $x \in H^+$. Then the following hold:
\begin{enumerate}
\item[$(i)$] If $x$ sees vertex $v_i$ for some index $i$, then $x$ sees $v_{i-2}$ and $v_{i+2}$.
\item[$(ii)$] $k$ is even, and either $N_H(x) = \{v_1,v_3,\ldots,v_{k-1}\}$ or 
$N_H(x) = \{v_2,v_4,\ldots,v_k\}$.
\end{enumerate}
\end{lemm}

\noindent
{\bf Proof.} (i): Assume without loss of generality that $x$ sees $v_1$. First, we show that $x$ sees either $v_3$ or $v_{k-1}$.
Assume to the contrary that $x$ sees neither $v_3$ nor $v_{k-1}$. Then to avoid that $x,v_1,v_3,v_{k-1},v_4$ induce a dart, $x$ sees $v_4$, and then by Proposition \ref{prop:1}, $x$ misses $v_k$ (since $x \in H^+$). Similarly by symmetry one has that $x$ sees $v_{k-2}$, and $x$ misses $v_2$. If $k = 7$, then $v_4$ misses $v_{k-2}$, a contradiction to Proposition \ref{prop:1}. If $k > 7$, then $v_4$ sees $v_{k-2}$, and then $v_{k-1},v_4,x,v_k,v_{k-2}$ induce a dart, a contradiction. Then $x$ sees either $v_3$ or $v_{k-1}$.

Then let us assume without loss of generality that $x$ sees $v_3$, so by Proposition \ref{prop:1}, $x$ misses $v_4$ and $v_k$ but then $x$ sees $v_{k-1}$ too, otherwise $v_k,v_3,x,v_1,v_{k-1}$ induce a dart.

\medskip

\noindent
(ii): By Proposition \ref{prop:1}, no vertex of $H^+$ dominates $H$. Thus, if $k$ is odd, then by an iterated application of statement (i), one has that each vertex $x \in H^+$ dominates $H$, a contradiction to Proposition \ref{prop:1}; if $k$ is even then by statement (i) the last part of statement (ii) follows.      
\qed

\medskip

\begin{coro}\label{dartfreenearlycoC7free}
Dart-free graphs are nearly $\overline{C_{2k+1}}$-free for $k \ge 3$.
\end{coro}

Theorem \ref{SPGT}, Lemma \ref{lemm:1} and Corollary \ref{dartfreenearlycoC7free} imply the following result:

\begin{theo}
$($Odd-hole,dart$)$-free graphs are nearly perfect.
\end{theo}

\noindent
{\bf Proof.} Let $G$ be an (odd hole, dart)-free graph. To prove the assertion it is sufficient to recall Theorem \ref{SPGT} and to 
show that if $G$ has a odd anti-hole $H$ with at least 7 vertices, then $H^+ = \emptyset$ which follows by contradiction from Lemma \ref{lemm:1} (ii).  
\qed

\medskip

Theorem \ref{basicMWISprop} (ii) implies:

\begin{coro}
The MWIS problem is solvable in polynomial time for $($odd-hole,dart$)$-free graphs.
\end{coro}

\subsection{Structure and MWIS for (Hole,Dart)-Free Graphs}

For (hole,dart)-free graphs, some stronger properties can be shown if one additionally excludes clique cutsets: 

\begin{lemm}\label{lemm:co-c7}
$($Hole, dart$)$-free atoms are nearly $\overline{C_k}$-free for $k \geq 7$.
\end{lemm}

\noindent
{\bf Proof.} Assume to the contrary that there is a vertex $v$ in $G$ such that the anti-neighborhood $A(v)$ of $v$ contains an induced $\overline{C_k}$, say $H$, for $k \geq 7$, with vertices $v_1,\ldots,v_k$ and co-edges $v_iv_{i+1}$ for $i = 1,\ldots,k$ (index arithmetic modulo $k$). Clearly $H^+ \neq \emptyset$ and $H^+$ is a cutset for $G$. Let $Q_v$ be the connected component of the anti-neighborhood $A(H)$ of $H$ containing $v$. Then since $G$ is an atom, there exist two vertices, say $x,y \in H^+$, such that: 
\begin{itemize}
\item[(a)] $x$ misses $y$, and 
\item[(b)] both $x$ and $y$ see some vertex of $Q_v$. 
\end{itemize}

Then by Lemma \ref{lemm:1}, let us distinguish between the following two cases, which are exhaustive by symmetry:

\begin{enumerate}
\item If $N_H(x) = \{v_1,v_3,\ldots,v_{k-1}\}$ and $N_H(y) = \{v_1,v_3,\ldots,v_{k-1}\}$, then $x,y,v_1,v_3,v_4$ induce a dart, a contradiction.
\item If $N_H(x) = \{v_1,v_3,\ldots,v_{k-1}\}$ and $N_H(y) = \{v_2,v_4,\ldots,v_{k}\}$, then $x,v_5,v_2,y$ induce a $P_4$; on the other hand, let $P$ be a shortest path form $x$ to $y$ in $G[Q_v \cup \{x,y\}]$; then the subgraph induced by $P,x,y,v_1,v_2$ is a hole, a contradiction.   
\end{enumerate}
This finally shows Lemma \ref{lemm:co-c7}.
\qed

\medskip

The proof of the subsequent Lemma \ref{lemm:co-c6} is similar to the one of Theorem 2 in \cite{BraGia2012}; in particular, the parts which are exactly the same (i.e., those which did not require the assumption co-chair-freeness) are reported in their original form.

\begin{lemm}\label{lemm:co-c6}
$($Hole, dart$)$-free atoms are nearly $\overline{C_6}$-free.
\end{lemm}

\noindent
{\bf Proof.}
Assume to the contrary that there is a vertex $v$ in $G$ such that its anti-neighborhood $A(v)$ contains an induced $\overline{C_6}$, say $A$, with vertices $v_1,\ldots,v_6$ such that $v_1,v_2,v_3$ is a clique left($A$), $v_4,v_5,v_6$ is a clique right($A$), and $v_1v_4, v_2v_5,$ and $v_3v_6$ are the edges between left($A$) and right($A$) (the {\em matching edges of A}). Let $A_i$ denote the neighbors of $A$ which see exactly $i$ vertices in $A$, $1 \leq i \leq 6$, and let $A^+$ denote the neighbors of $A$ which see a vertex in the connected component $Q_v$ of the anti-neighborhood of $A$ containing $v$. Note that $A^+$ depends on $v$ but for short (and in order to avoid confusion with anti-neighborhoods) we write $A^+$ instead of $A^+(v)$.

Let also $A_2^+(1,4)$ denote the vertices of $A_2^+$ which see exactly $v_1$ and $v_4$ in $A$ and similarly for some other cases. We first collect some simple properties.

\begin{clai}\label{clai:3}
$A_4^+ = A_5^+ = A_6^+ = \emptyset$
\end{clai}

\noindent
{\em Proof of} Claim \ref{clai:3}. Follows by Proposition \ref{prop:1}. 
${\diamond}$

\begin{clai}\label{clai:4}
If $x \in A_2$ then $x$ sees the two vertices of a matching edge in $A$.
\end{clai}

\noindent
{\em Proof of} Claim \ref{clai:4}. Assume not; then $x$ is either adjacent to two vertices in left($A$) (right($A$) respectively), say $x$ sees $v_1,v_2$, in which case $v_4,v_3,v_2,v_1,x$ induce a dart, or $x$ is adjacent to two nonadjacent vertices, say $x$ sees $v_1,v_5$ in which case $x,v_1,v_3,v_6,v_5$ induce a $C_5$. ${\diamond}$

\begin{clai}\label{clai:5}
If $x \in A_3^+$ then $x$ has a join to $left(A)$ or $x$ has a join to $right(A)$.
\end{clai}

\noindent
{\em Proof of} Claim \ref{clai:5}. Let us consider the following cases which are exhaustive by symmetry. If $x$ sees three vertices of $A$ inducing a $P_3$, then one has a contradiction to Proposition \ref{prop:1}. If $x$ sees two vertices of left($A$), say $v_1,v_2$, and a vertex of right($A$) missing $v_1,v_2$, that is vertex $v_6$, then $x,v_1,v_2,v_3,v_4$ induce a dart. 
${\diamond}$

\medskip

Since $G$ is hole-free, we have:

\begin{clai}\label{clai:7}
For all $x,y \in A^+$ with $x$ missing $y$, we have $N_A(x) \subseteq N_A(y)$ or vice versa. Moreover, $x$ and $y$ have a common neighbor in $Q_v$.
\end{clai}

\begin{clai}\label{clai:8}
If $A_2^+ \neq \emptyset$ then $A_3^+ = \emptyset$ and vice versa.
\end{clai}

\noindent
{\em Proof of} Claim \ref{clai:8}. Assume not; let $x \in A_2^+$ and $y \in A_3^+$, say $N_A(x) = \{v_1,v_4\}$ and $N_A(y) = \{v_1,v_2,v_3\}$ by Claims \ref{clai:4} and \ref{clai:5}. Then since $y,v_3,v_6,v_4,x$ induce no $C_5$, $x$ misses $y$ but now, by Claim \ref{clai:7}, the neighborhoods of $x$ and $y$ must be comparable - contradiction. 
${\diamond}$

\begin{clai}\label{clai:9}
At most one of $A_2^+(1,4),A_2^+(2,5),A_2^+(3,6)$ is nonempty.
\end{clai}

\noindent
{\em Proof of} Claim \ref{clai:9}. Assume not; without loss of generality, let $x \in A_2^+(1,4)$ and $y \in A_2^+(2,5)$. Then by Claim \ref{clai:7}, $x$ sees $y$ but now $x,y,v_2,v_3,v_6,v_4$ induce a $C_6$. 
${\diamond}$

\begin{clai}\label{clai:10}
The set $N_A(A_1^+)$ of neighbors of all $x \in A_1^+$ in $A$ is a clique.
\end{clai}

\noindent
{\em Proof of} Claim \ref{clai:10}. Assume not; let without loss of generality $x \in A_1^+$ see $v_1$ and $y \in A_1^+$ see $v_5$. Then since $x,v_1,v_2,v_5,y$ do not induce a $C_5$, $x$ misses $y$ but now by Claim \ref{clai:7}, $N_A(x)$ and $N_A(y)$ must be comparable - contradiction. 
${\diamond}$

\begin{clai}\label{clai:11}
No vertex in $A_1^+$ sees a vertex in $A_2^+ \cup A_3^+$, i.e., $A_1^+ {\0} A_2^+ \cup A_3^+$.
\end{clai}

\noindent
{\em Proof of} Claim \ref{clai:11}. Assume not; let $x \in A_1^+$ and first assume that $y \in A_2^+$, say $N_A(y) = \{v_1,v_4\}$ with $x$ seeing $y$. If $x$ sees $v_1$, then $x,v_1,y,v_4,v_2$ induce a dart, and if $x$ sees $v_2$ then $x,y,v_2,v_5,v_4$ induce a $C_5$. The other cases are symmetric. This shows Claim \ref{clai:11}. 
${\diamond}$

\begin{clai}\label{clai:12}
$A_3^+$ is a clique.
\end{clai}

\noindent
{\em Proof of} Claim \ref{clai:12}. First note that $A_3^+(1,2,3){\1}A_3^+(4,5,6)$ since $G$ is $C_5$-free. If there are $x,y \in A_3^+(1,2,3)$ with $x$ missing $y$ then $v_4,v_1,v_2,x,y$ induce a dart. Then $A_3^+(1,2,3)$ is a clique. The fact that $A_3^+(4,5,6)$ is a clique is shown analogously. 
${\diamond}$

\medskip

Now we conclude that in any case, we get a clique separator between $Q_v$ and some vertex in $A$ (which finally contradicts to the assumption that $G$ is an atom):

\medskip

\noindent
{\bf Case 1.} $A_3^+ \neq \emptyset$. 

\medskip

Then by Claim \ref{clai:8}, $A_2^+ = \emptyset$.
First suppose that $A_1^+ \neq \emptyset$. We claim that $N_A(A_1^+) \cup A_3^+$ is a clique separator: Recall that by Claim \ref{clai:10}, $N_A(A_1^+)$ is a clique, by Claim \ref{clai:12}, $A_3^+$ is a clique, and by Claims \ref{clai:7} and \ref{clai:11}, every $x \in A_3^+$ sees every $y \in N_A(A_1^+)$ (note that if $N_A(A_1^+) \subseteq \{v_1,v_2,v_3\}$ then $A_3^+(4,5,6) = \emptyset$, and the case that $N_A(A_1^+)$ is one of the matching edges is impossible if $A_3^+ \neq \emptyset$). Obviously, $N_A(A_1^+) \cup A_3^+$ is a separator between $v$ and a nonempty part of $A$.

Now suppose that $A_1^+ = \emptyset$. Then $A_3^+$ is a clique separator (in this case, possibly $A_3^+(1,2,3) \neq \emptyset$ and $A_3^+(4,5,6) \neq \emptyset$).

\medskip

\noindent
{\bf Case 2.} $A_2^+ \neq \emptyset$. 

\medskip

Then by Claim \ref{clai:8}, $A_3^+ = \emptyset$, and by Claim \ref{clai:9}, at most one of the sets $A_2^+(1,4)$, $A_2^+(2,5)$, $A_2^+(3,6)$ is nonempty, say $A_2^+(1,4) \neq \emptyset$ and $A_2^+(2,5) = A_2^+(3,6) = \emptyset$. Then $\{v_1,v_4\}$ is a clique separator (note that in this case, by Claims \ref{clai:7} and \ref{clai:11}, $N_A(A_1^+) \subseteq \{v_1,v_4\}$).

\medskip
\noindent
{\bf Case 3.} $A_2^+ \cup A_3^+ = \emptyset$. 

\medskip

Then $A_1^+ \neq \emptyset$ since $G$ is connected, and again by Claim \ref{clai:10}, $N_A(A_1^+)$ is a clique separator.
This finishes the proof of Lemma \ref{lemm:co-c6}.             
\qed

\medskip

By Lemmas \ref{lemm:co-c7} and \ref{lemm:co-c6}, one obtains the following result:

\begin{theo}\label{holedartnearlywc}
$($Hole, dart$)$-free atoms are nearly weakly chordal.
\end{theo}

By Theorem \ref{basicMWISprop} (i), we obtain:

\begin{coro}
The MWIS problem is solvable in polynomial time for $($hole, dart$)$-free graphs.
\end{coro}

The corresponding result in \cite{BasChaKar2012} gives a time bound of ${\cal O}(n^4)$ for the MWIS problem on $($hole, dart$)$-free graphs. This result is based on Theorem 4 in \cite{BasChaKar2012} showing that (hole,dart)-free atoms are nearly gem-free. Their time bound is better than the obvious time bound resulting from Theorem \ref{holedartnearlywc} and Theorem \ref{basicMWISprop} (ii). However, more can be said about the structure of (hole,dart)-free graphs: Lemma 2 in \cite{Brand2004} implies that prime dart- and gem-free graphs are diamond-free. This is slightly better than Lemma 1 in \cite{BasChaKar2012} saying that prime dart- and gem-free graphs are paraglider-free.  

Moreover, we need the following result:

\begin{theo}[\cite{BerBraGiaMaf2012}]\label{holediamondcoC6freeatoms}   
$($Hole,diamond$)$-free atoms that contain no induced $\overline{C_6}$ are either a clique or chordal bipartite. 
\end{theo}

Theorem \ref{holediamondcoC6freeatoms} and Lemma \ref{lemm:co-c6} imply: 

\begin{coro}\label{holedartfrstructure}
$($Hole, dart$)$-free prime atoms are either nearly a clique or nearly chordal bipartite. 
\end{coro}

This does not improve the ${\cal O}(n^4)$ time bound of \cite{BasChaKar2012} but gives more structural insight as asked for in the last paragraph of \cite{BasChaKar2012}.  
 
\section{Structure and MWIS for (Odd-)Hole- and Bull-Free Graphs}\label{holebullfr}

In this section, we show that prime odd-hole- and bull-free graphs are nearly perfect and that prime hole- and bull-free atoms are nearly weakly chordal. This is the main result of the section and implies polynomial time for MWIS on both graph classes (with better time bound for (hole, bull)-free graphs). We first collect some properties of prime bull-free graphs. 

\medskip

Let $G$ be a prime graph with at least 7 vertices, and suppose that $G$ contains a $\overline{C_k}$, say $H$, for some $k \ge 6$. Since $G$ is prime, $H$ is not a module, and thus, there is a vertex $z \notin V(H)$ distinguishing vertices $x,y \in V(H)$, say $xz \in E$ and $yz \notin E$. Since $H$ is connected, we can assume without loss of generality that $z$ distinguishes an edge $xy \in E$ in $H$. Let $H_0:=H$ and for $k \ge 1$, let $H_k$ result by adding a distinguishing vertex $z_k \notin V(H_{k-1})$ to $H_{k-1}$, that is, $H_k:=G[V(H_{k-1}) \cup \{z_k\}]$. As before, we can assume without loss of generality that $z_k$ distinguishes an edge in $H_{k-1}$. This defines a strictly increasing sequence of induced subgraphs of $G$. Since $G$ is finite, there is a largest $\hat{k}$ such that $H_{\hat{k}}$ exists.    

\begin{lemm}\label{lemm:bull}
If $G$ is a connected bull-free graph containing some induced subgraph $H_k$, $k \ge 0$, as defined above, with nonempty anti-neighborhood $A(H_k)$, and $u \in H_k^+$ then $u$ has a join to $H_k$. 
\end{lemm}

\noindent 
{\bf Proof.}
We show Lemma \ref{lemm:bull} by induction on $k$. For $k=0$, the proof goes as follows. Let $u \in H_0^+$ and let $v \in A(H_0)$ be a neighbor of $u$. To prove the assertion, let us show that if $u$ sees a vertex $v_i$ for some index $i$, then $u$ sees also $v_{i-1}$ and $v_{i+1}$ (index arithmetic modulo $k$); by iterating this argument the assertion follows.

Then let us assume without loss of generality that $u$ sees $v_1$, and show that $u$ sees $v_k$ and $v_2$. To avoid that $u,v_1,v_3,v_{k-1},v_k$ induce a bull, $u$ sees either $v_3$ or $v_{k-1}$ or $v_k$. Moreover, by a symmetry argument, one has that $u$ sees either $v_3$ or $v_{k-1}$ or $v_2$. Also, if $u$ sees $v_k$ and $v_2$, then $u$ sees $v_{k-1}$ and $v_3$, otherwise $v,u,v_k,v_2,v_{k-1}$ or $v,u,v_k,v_2,v_3$ induce a bull. All the previous facts imply that $u$ sees either $v_{k-1}$ or $v_3$; without loss of generality, say $u$ sees $v_3$. Then $u$ sees $v_k$, otherwise $v,u,v_1,v_3,v_k$ induce a bull. Then $u$ sees $v_2$, otherwise $v,u,v_3,v_k,v_2$ induce a bull. This completes the proof for $k=0$. 

\medskip

Now, as the induction hypothesis, assume that the assertion is true for values at most $k$, and let us show that it holds for $k+1$. Let $z \notin V(H_k)$ be any vertex distinguishing an edge $xy$ in $H_k$, say $xz \in E$ and $yz \notin E$, and let $H_{k+1}=G[V(H_k) \cup \{z\}]$. 
Suppose that $A(H_{k+1}) \neq \emptyset$, and without loss of generality choose a vertex $v \in A(H_{k+1})$ such that the distance between $v$ and $H_{k+1}$ is 2. Now, also $v \in A(H_{k})$ holds, and the distance between $v$ and $H_k$ is at least 2. If the distance is 2 then let $u \in H_k^+$ be a neighbor of $v$. Thus, $v$ misses $H_k$ and, by the induction hypothesis, any neighbor $u \in H_k^+$ of $v$ has a join to $H_k$. Since $v \in A(H_{k+1})$, $vz \notin E$ holds. Since $G$ is bull-free, $v,u,x,y,z$ is not a bull, and since $v$ misses $x,y,z$ and $u$ sees $x,y$, it follows that $uz \in E$ and thus, $u$ has a join to $H_{k+1}$. 

In the other case, assume that the distance between $v$ and $H_k$ is at least 3, and recall that the distance between $v$ and $H_{k+1}$ is 2; thus, let $u$ be a common neighbor of $v$ and $z$. Note that in this case, $u$ misses $H_k$, and now by the induction hypothesis, $z \in H_k^+$ must have a join to $H_k$ which is a contradiction.  
This shows Lemma \ref{lemm:bull}. 
\qed

\begin{lemm}\label{lemm:bullnearlycoCkfr}
Prime bull-free graphs are nearly $\overline{C_\ell}$-free for any $\ell \ge 6$.
\end{lemm}

\noindent 
{\bf Proof.} Suppose that there is a vertex $v$ such that $A(v)$ contains a $\overline{C_\ell}$, say $H$, for some $\ell \ge 6$. Without loss of generality, let $v$ be in distance 2 to $H$, and let $u \in H^+$. Thus, by definition and by Lemma \ref{lemm:bull}, $H$ is contained in $A(v) \cap N(u)$. Now let $H_k$, $k \ge 0$, and $H_{\hat{k}}$ be defined as above. We claim that $v$ misses $H_{\hat{k}}$: Clearly $v$ misses $H_0$. Suppose that $v$ misses $H_k$ and sees $H_{k+1}$; recall that $H_{k+1}$ results by adding a distinguishing vertex $z_k$ to $H_k$; say, for some edge $xy$ in $H_k$, $z_k$ sees $x$ and misses $y$. Now if $v$ misses $H_k$ and sees $H_{k+1}$ then $vz_k \in E$, and now, since $z_k \in H_k^+$, by Lemma \ref{lemm:bull}, $z_k$ should be universal for $H_k$ - a contradiction.     

Thus, by Lemma \ref{lemm:bull}, it follows that $H_{\hat{k}}$ is a module in $G$ which is a contradiction to the assumption that $G$ is prime.   
\qed

\medskip

By the Strong Perfect Graph Theorem and by the definition of weakly chordal graphs, it follows:

\begin{coro}\label{holebullfrnearlywc}
Prime $($odd-hole,bull$)$-free graphs are nearly perfect, and prime $($hole,bull$)$-free graphs are nearly weakly chordal.
\end{coro}

By Theorem \ref{MWISprimered}, Corollary \ref{nearlyPiMWIS} and Theorem \ref{basicMWISprop}, we obtain: 

\begin{coro}\label{holebullfrMWIS}
For $($odd-hole,bull$)$-free graphs, the MWIS problem is solvable in polynomial time. 
\end{coro}

The time bound for (hole,bull)-free graphs is ${\cal O}(n^5)$.

\medskip

Note that Corollary \ref{holebullfrMWIS} concerning (odd-hole,bull)-free graphs is close to the MWIS result implied by the structure result of De Simone \cite{DeSim1993}, i.e., MWIS is polynomial for graphs with no odd apples and no six specific induced subgraphs, five of which contain a bull. It is also well known that MWIS can be solved in polynomial time for $P_5$- and bull-free graphs since such graphs do not contain any of the forbidden subgraphs in \cite{DeSim1993}.
However, for $P_5$- and bull-free graphs, we can say more; the proof of Lemma \ref{lemm:bull} can be adapted to similar cases as we describe subsequently.   

\section{MWIS for $P_5$- and Bull-Free Graphs Revisited}\label{P5bullfr}

\begin{lemm}\label{lemm:bullnearlyC5housefr}
Prime $P_5$- and bull-free graphs are nearly $C_5$-free and nearly house-free.
\end{lemm}
 
\noindent 
{\bf Proof.}
We recursively define a sequence $H_k$, $k \ge 0$, of increasing subgraphs for which $A(H_k)$ is nonempty, $v \in A(H_k)$ and $u \in H^+_k$ being a neighbor of $v$ in the same way as before Lemma \ref{lemm:bull}; only $H_0$ is different. For showing that prime $P_5$- and bull-free graphs are nearly $C_5$-free, $H_0$ is a $C_5$, say with vertices $v_1,\ldots,v_5$ and edges $v_iv_{i+1}$ (index arithmetic modulo 5), and for showing that prime $P_5$- and bull-free graphs are nearly house-free, $H_0$ is a house, say with vertices $v_1,\ldots,v_5$ such that $v_1,v_2,v_3,v_4$ induce a $C_4$ and $v_5$ sees $v_2$ and $v_3$. 
  
We show Lemma \ref{lemm:bullnearlyC5housefr} by induction on $k$. Let $G$ be a prime ($P_5$,bull)-free graph. For $k=0$, we first consider the case when $H_0$ is a $C_5$. We need the following notion: 

For a subgraph $H$ of $G$, a vertex $x \notin V(H)$ is an {\em $i$-vertex of $H$} if $x$ sees exactly $i$ vertices in $H$. Obviously, $C_5$ has no 1-vertex since $G$ is $P_5$-free, and $C_5$ has no 2-vertex since two consecutive neighbors lead to a bull, and two nonconsecutive neighbors lead to $P_5$. Similarly, if for a 3-vertex, not all neighbors are consecutive, we get a bull, and for consecutive neighbors, we get a $P_5$. Any 4-vertex leads to a bull. Thus, any vertex outside $H_0=C_5$ seeing $H_0$ is universal for $H_0$. Now with the same inductive arguments as in the proof of Lemma \ref{lemm:bull},
we show that any vertex $u \in H_k^+$ has a join to $H_k$. With the same arguments as in Lemma \ref{lemm:bullnearlycoCkfr}, we obtain that $G$ is nearly $C_5$-free. 

\medskip
Now, let $H_0$ be a house as described above.  

\begin{enumerate}
\item For a 1-vertex $u$, we obtain a $P_5$ if $u$ is adjacent to $v_1$ or $v_4$ or a bull if $u$ is adjacent to $v_2,v_3$ or $v_5$. 

\item For a 2-vertex $u$, there are two cases: 

\begin{enumerate}
\item $u$ has two neighbors in the $C_4$:  
\begin{enumerate}
\item if $u$ sees $v_1$ and $v_2$ then $u,v_1,v_2,v_4,v_5$ induce a bull, 
\item if $u$ sees $v_1$ and $v_3$ then $v,u,v_1,v_2,v_5$ induce a $P_5$,      
\item if $u$ sees $v_2$ and $v_3$ then $v,u,v_2,v_3,v_4$ induce a bull, and
\item if $u$ sees $v_1$ and $v_4$ then $v,u,v_1,v_3,v_4$ induce a bull. 
\end{enumerate}

\item $u$ has one neighbor in the $C_4$ and sees $v_5$:
\begin{enumerate}
\item if $u$ sees $v_1$ then $u,v_2,v_3,v_4,v_5$ induce a bull, and 
\item if $u$ sees $v_2$ then $v,u,v_1,v_2,v_5$ induce a bull.      
\end{enumerate}
\end{enumerate}

\item For a 3-vertex $u$, there are two cases: 

\begin{enumerate}
\item $u$ has three neighbors in the $C_4$:  
\begin{enumerate}
\item if $u$ sees $v_1$, $v_2$ and $v_3$ then $u,v_1,v_2,v_4,v_5$ induce a bull, and 
\item if $u$ sees $v_1$, $v_2$ and $v_4$ then $v,u,v_1,v_3,v_4$ induce a bull.      
\end{enumerate}

\item $u$ has two neighbors in the $C_4$ and sees $v_5$:
\begin{enumerate}
\item if $u$ sees $v_1$ and $v_2$ then $v,u,v_1,v_2,v_4$ induce a bull, 
\item if $u$ sees $v_1$ and $v_3$ then $v,u,v_3,v_4,v_5$ induce a bull,      
\item if $u$ sees $v_2$ and $v_3$ then $v,u,v_2,v_3,v_4$ induce a bull, and 
\item if $u$ sees $v_1$ and $v_4$ then $v,u,v_1,v_3,v_4$ induce a bull.          
\end{enumerate}
\end{enumerate}

\item For a 4-vertex $u$, there are two cases: 

\begin{enumerate}
\item If $u$ has four neighbors in the $C_4$ then $v,u,v_3,v_4,v_5$ induce a bull. 

\item Otherwise, if $u$ has three neighbors in the $C_4$ and sees $v_5$ then:
\begin{enumerate}
\item if $u$ sees $v_1$, $v_2$ and $v_3$ then $v,u,v_1,v_2,v_4$ induce a bull, and  
\item if $u$ sees $v_1$, $v_2$ and $v_4$ then $v,u,v_1,v_3,v_4$ induce a bull.      
\end{enumerate}
\end{enumerate} 
 
\end{enumerate}
 
Thus, $u$ is universal for $H_0$, and with the same arguments as above, we obtain that $G$ is nearly house-free.
\qed  

\medskip

A bipartite graph $B$ with color classes $X$ and $Y$ is a {\em bipartite chain graph} if the neighborhoods of the vertices of one color class form an increasing sequence with respect to set inclusion. It is well known that these graphs can be recognized in linear time, have bounded clique-width, and MWIS can be solved in linear time for them. From a result by Fouquet \cite{Fouqu1993}, it follows: 
 
\begin{coro}\label{P5bullfrnearlychain}
Prime $(P_5$,bull$)$-free graphs are nearly bipartite chain graphs or nearly co-bipartite chain graphs. 
\end{coro}

\begin{coro}\label{P5bullfrMWIS}
For $(P_5$,bull$)$-free graphs, the MWIS problem is solvable in time ${\cal O}(n m)$. 
\end{coro}

A polynomial time algorithm for MWIS on ($P_5$,dart)-free graphs follows from \cite{Mosca2004}. 

\section{Conclusion}

The class of hole-free graphs is closely related to many well-studied graphs classes, such as chordal graphs, weakly chordal graphs and perfect graphs. However the complexity of the MWIS problem for hole-free graphs is open. Following a recent line of research \cite{BraGia2012,BraGiaMaf2012}, we show that MWIS can be solved in polynomial time for odd-hole- and dart-free graphs as well as for odd-hole- and bull-free graphs. For this purpose, the main structural results of this paper are the following (which might also be useful in other contexts):  

\begin{enumerate}
\item 
\begin{enumerate}
\item Dart-free graphs are nearly $\overline{C_{2k+1}}$-free for $k \ge 3$. 
\item Hole- and dart-free atoms are nearly $\overline{C_k}$-free for $k \geq 6$.
\item Consequently, odd-hole- and dart-free graphs are nearly perfect, and hole- and dart-free atoms are nearly weakly chordal.  
\end{enumerate}

\item 
\begin{enumerate}
\item Prime bull-free graphs are nearly $\overline{C_k}$-free for any $k \ge 6$.
\item Consequently, prime odd-hole- and bull-free graphs are nearly perfect, and prime hole- and bull-free atoms are nearly weakly chordal. 
\end{enumerate}

\item Using the approach for bull-free graphs, we show that prime ($P_5$,bull)-free graphs are nearly bipartite chain graphs or nearly co-bipartite chain graphs; this leads to a better time bound for MWIS on ($P_5$,bull)-free graphs.  
\end{enumerate}

The results on dart-free graphs imply that MWIS is solvable in polynomial time for odd-hole- and dart-free graphs, by finally reducing the problem to perfect graphs. Note that (hole, dart)-free graphs can be recognized in polynomial time since recognition of hole-free graphs can be done in polynomial time by \cite{Spinrad1991} (see also \cite{ChvFonSunZem2002} for recognizing dart-free perfect graphs in polynomial time, though in general perfect graphs can be recognized in polynomial time by \cite{ChuCorLiuSeyVus2005,CorLiuVus2003}).

\medskip

The results on bull-free graphs allow to solve MWIS in polynomial time for odd-hole- and bull-free graphs, again by finally reducing the problem to perfect graphs. Note that (hole, bull)-free graphs can be recognized in polynomial time since recognition of hole-free graphs can be done in polynomial time by \cite{Spinrad1991}. 

\medskip

Let us conclude by a remark on the class of (hole,gem)-free graphs. It seems that the situation for (hole,gem)-free graphs is more complicated than for (hole,dart)-free and (hole,bull)-free graphs. Clearly, by the Strong Perfect Graph Theorem, (hole,gem)-free graphs are perfect, and thus, the MWIS problem is solvable in polynomial time for (hole,gem)-free graphs. However, we would like to find a direct combinatorial algorithm with a good time bound as in the other cases. Therefore we conclude with the following: 

\medskip

\noindent
{\bf Open Problem.} What is a good time bound for the MWIS problem for (hole,gem)-free graphs?



\begin{footnotesize}
\renewcommand{\baselinestretch}{0.4}

\end{footnotesize}

\end{document}